\documentclass{article}
\usepackage{graphicx,amsmath,amssymb}
\def\A{{\cal A}}

\def\F{{\cal F}}

\def\K{{\cal K}}

\def\M{{\cal M}}
\def\N{{\cal N}}

\def\linebreak{\hfill\break}

\def\bra<#1|{\langle #1\rvert}
\def\ket|#1>{\lvert#1 \rangle}
\def\braket<#1|#2>{\langle #1|#2 \rangle}

\def\then{\Rightarrow\quad}

\def\therefore{\mbox{\setbox0=\hbox{X}\hbox{$\ldotp$}\raise0.7\ht0\hbox{$\ldotp$}\hbox{$\ldotp$}} \quad }
\def\because{\mbox{\setbox0=\hbox{X}\raise0.7\ht0\hbox{$\ldotp$}\hbox{$\ldotp$}\raise0.7\ht0\hbox{$\ldotp$}}\kern0pt }

\def\maps{\rightarrow}

\def\Frac(#1/#2){\left(\frac{#1}{#2}\right)}

\def\Eq#1{\begin{equation} #1 \end{equation}}

\def\Eqr#1{\begin{eqnarray} #1 \end{eqnarray}}

\def\Bitm{\begin{itemize}}
\def\Eitm{\end{itemize}}
\def\Blist#1#2{\begin{list}{#1}{\parsep=0pt \itemsep=0pt%
  \listparindent=0pt #2}}
\def\Elist{\end{list}}
\def\ignore#1#2{\def\ignoreflag{#1}\long\def\tmptext{#2}
  \ifnum\ignoreflag>1 #2 \fi}

\def\THB{{\mathbb T}}
\def\VHB{{\mathbb V}}
\def\SHB{{\mathbb S}}

\begin{document}
\title{Stability of Generalised Static Black Holes\\ 
in Higher Dimensions%
\footnote{A talk given at the Seventh Hungarian Relativity Workshop, 10--15 Aug. 2003, Sarospatak, Hungary.}}

\author{{\it Hideo Kodama$^1$ and Akihiro Ishibashi$^2$}\\
$^1$Yukawa Institute for Theoretical Physics\\ 
Kyoto University, Kyoto 606-8502 (Japan) \\
E-mail: kodama@yukawa.kyoto-u.ac.jp\\ 
$^2$D.A.M.T.P., Centre for Mathematical Sciences\\ 
University of Cambridge, Wilberforce Road\\ 
Cambridge CB3 0WA (UK)\\ 
E-mail: A.Ishibashi@damtp.cam.ac.uk}

\maketitle

\abstract{%
We discuss the stability of (charged) static black holes in 
higher-dimensional spacetimes with and without cosmological constant 
by using gauge-invariant master equations of the Schr\"odinger 
equation type for black hole perturbations derived by the authors 
recently. In particular, we show that the stability of 
higher-dimensional Schwarzschild black holes can be proved with the 
help of a technique called {\em S-deformation} of the master 
equations. We also point out that higher-dimensional static black 
holes might be unstable only against scalar-type perturbations in 
the neutral case and in the charged case with spherically symmetric 
or flat horizons. }

\section{Introduction}  \label{intro}

The perturbation analysis of 4-dimensional black holes has 
a long history and has provided valuable information on 
various problems such as the fate of gravitational 
collapse, the stability and uniqueness of black holes, 
cosmic censorship, and gravitational wave emissions. From 
a technical point of view, the key point in this 
perturbation analysis was the fact that the perturbation 
equations can be reduced to a single 2nd-order ordinary 
differential equation (ODE) for a master variable $\Phi$ 
of the form
\Eq{
-\frac{d^2 \Phi}{dr_*^2}+\left[V(r)-\omega^2  \right]\Phi=S.
}

Recently, starting from a gauge-invariant formalism for 
perturbations of higher-dimensional spacetimes with 
spatial symmetry\cite{Kodama.H&Ishibashi&Seto2000}, we 
showed that similar reduction to single master equations 
is also possible for perturbations of generalised static 
black holes with or without charge in higher dimensions%
\cite{Kodama.H&Ishibashi2003A,Kodama.H&Ishibashi2003Aa}.
Further, using these master equations, we analysed the 
stability of such black holes%
\cite{Ishibashi.A&Kodama2003A,Kodama.H&Ishibashi2003Aa}.  

In this talk, we explain the basic idea of this formulation 
and describe the results of the stability analysis of 
(charged) static black holes in higher dimensions.

\section{Tensorial Decomposition of Perturbations}

\subsection{Unperturbed Background}

We assume that the unperturbed background spacetime is 
locally written $ \M^D\approx \N^2\times \K^n$ and its 
metric has the form
\Eq{
ds^2=g_{ab}(y)dy^a dy^b + r^2(y)d\sigma_n^2,
\label{BG:metric}}
where $d\sigma_n^2=\gamma_{ij}(z)dz^i dz^j$ is a metric 
of an $n$-dimensional complete Einstein space $\K^n$ with $\hat 
R_{ij}=(n-1)K \gamma_{ij}$ ($K=0,\pm1$), and $g_{ab}$ is a static 
metric of the 2-dimensional space $\N^2$ expressed as
\Eq{
g_{ab}dy^ady^b= -f(r)dt^2+\frac{dr^2}{f(r)}.
\label{StaticOrbitSpace}
}
%
In this and the next section, we assume that this metric is a 
solution to the vacuum Einstein equations with cosmological constant 
$\Lambda=n(n+1)\lambda/2$. Hence, $f(r)$ is expressed 
as\cite{Birmingham.D1999}
\Eq{
f(r)=K - \frac{2M}{r^{n-1}} -\lambda r^2.
\label{f:neutral}
}

When the spacetime contains a regular black hole, the 
space $\K^n$ represents a spatial section of the horizon. 
For $K=1$, $\K^n$ is always compact from Meyers' theorem and for $n\le3$, $\K^n$ is locally isomorphic to $S^n$. The spacetime contains a regular black hole if $\lambda M^{2/(n-1)}<(n-1)/(n+1)^{(n+1)/(n-1)}$. 
In contrast, for $K=0$ or $-1$, $\K^n$ may not be compact, 
and the spacetime contains a regular black hole only for 
$\lambda<0$.

\subsection{Perturbations}


In terms of the perturbation variable $\psi_{\mu\nu}=h_{\mu\nu}-h 
g_{\mu\nu}/2$ with $h_{\mu\nu}=\delta g_{\mu\nu}$, the perturbed 
vacuum Einstein equations are written as
\Eq{
-\nabla^2 \psi_{\mu\nu} -2R_{\mu\alpha\nu\beta}\psi^{\alpha\beta}
  +2\nabla_{(\mu}\nabla^\alpha \psi_{\nu)\beta}
  -\nabla^\alpha\nabla^\beta\psi_{\alpha\beta} g_{\mu\nu}=0.
\label{PerturbationEq:general}}
In order to analyse the behaviour of perturbations using this equation, we must solve two problems. First, this equation is invariant under the gauge transformation
\Eq{
x^\mu \maps x^\mu + \xi^\mu\quad \then \quad
\bar\delta h_{\mu\nu}= -\nabla_\mu\xi_\nu -\nabla_\nu\xi_\mu.
}
In order to extract the dynamics of the physical degrees of freedom, 
this gauge freedom should be eliminated. Second,  
\eqref{PerturbationEq:general} is actually a set of coupled 
equations with $(n+2)(n+3)/2$ entries and is quite hard to analyse 
in general.

For the special background \eqref{BG:metric}, the second problem can 
be made tractable with the help of the following tensorial 
decomposition of $h_{\mu\nu}$. First, according to the tensorial 
behaviour on $\K^n$, $h_{\mu\nu}$ can be divided into scalars 
$h_{ab}$, vectors $h_{ai}$ and a tensor $h_{ij}$. The vector and 
tensor components can be further decomposed as%
\cite{Kodama.H&Sasaki1984,Kodama.H&Ishibashi2003Aa}
\Eqr{
&& h_{ai}=\hat D_i h_a + h^{(1)}_{ai};\quad \hat D^ih^{(1)}_{ai}=0,\\
&& h_{ij}=h_L\gamma_{ij}+h_{Tij}; \quad h^j_{Tj}=0,\\
&& h_{Tij}=\left(  \hat D_i\hat D_j
      -\frac{1}{n}\gamma_{ij}\hat\triangle\right)h^{(0)}_{T}
   +2\hat D_{(i}h^{(1)}_{T j)}+h^{(2)}_{T ij};\notag\\
&&   \hat D^j h^{(1)}_{T j}=0,\ 
   \hat D^j h^{(2)}_{T ij}=0,
}
where $\hat D_i$ is the covariant derivative with respect to the 
metric $\gamma_{ij}$ on $\K^n$ and $\hat\triangle=\hat D\cdot\hat 
D$. By this decomposition, we obtain the following three groups of 
variables:
\Blist{$\bullet$}{}
\item { Scalar-type} variables:  $h_{ab}$, $h_a$, $h_L$, $h_T^{(0)}$,
\item { Vector-type} variables:  $h^{(1)}_{ai}$, $h^{(1)}_{Ti}$,
\item { Tensor-type} variable:   $h^{(2)}_{Tij}$.
\Elist
The Einstein equations written in terms of these variables are 
divided into three subsets each of which contains only variables 
belonging to one of the above three sets of variables. Further, the 
tensorial indices for vector-type and tensor-type variables can be 
eliminated if we expand these variables in terms of harmonic 
tensors on $\K^n$, and gauge-invariant variables can be easily 
constructed from the harmonic expansion coefficients, as we will 
show below. In this way, the Einstein equations for perturbations of 
each type can be reduced to a set of gauge-invariant equations with 
a small and $n$-independent number of 
entries\cite{Kodama.H&Ishibashi&Seto2000}. In particular, for tensor 
perturbations, this procedure gives a single wave equation for a 
single variable.

\section{Stability Analysis}

\subsection{4D Schwarzschild black hole}

In a four-dimensional background spacetime with a Schwarzschild 
black hole, it was shown by Regge and 
Wheeler\cite{Regge.T&Wheeler1957} and Zerilli\cite{Zerilli.F1970} 
that by the harmonic expansion on $\K^2=S^2$ and the Fourier 
transformation with respect to time,  $h_{ab}=f_{ab}(r)e^{-i\omega 
t} Y^m_l(\theta,\phi), \cdots$, the perturbation equations for any 
type of perturbations can be eventually reduced to a single 
Schr\"odinger-type second-order ODE 
\Eq{
 -\frac{d^2 \Phi}{dr_*^2}+ V(r)\Phi = \omega^2 \Phi,
}
where $r_*= \int^r dr/f(r)$ and the master variable $\Phi$ is a 
combination of the harmonic expansion coefficients $h_{ab}, \cdots$. 
The effective potential $V=V_V$ for vector perturbations (axial 
modes) is given by the Regge-Wheeler potential
\Eq{
V_V=\frac{f}{r^2}\left( m+2-\frac{6M}{r} \right),
}
and the effective potential $V=V_S$ for scalar perturbations (polar 
modes) is given by the Zerilli potential
\Eq{
V_S=\frac{f}{r^2H^2}\left( m^2(m+2)+ \frac{6m^2M}{r}
      +\frac{36mM^2}{r^2}+\frac{72M^3}{r^3} \right),
}
where $m=(l-1)(l+2)$($l=2,3,\cdots$) and $H=m+6M/r$. A tensor-type perturbation does not exist for $n=2$.

Since the effective potentials $V_V$ and $V_S$ are positive outside 
the horizon $(r>2M)$, the master equations for $\Phi$ have no 
regular and normalizable solution for $\omega^2<0$. Hence, the 4D 
Schwarzschild black hole is perturbatively 
stable\cite{Vishveshwara.C1970}.

Similar master equations for 4D black holes with $\Lambda\not=0$ 
were derived by Cardoso and 
Lemos\cite{Cardoso.V&Lemos2001,Cardoso.V&Lemos2001a}. With the help 
of these equations, the stability of these black holes were proved 
by us in Ref.\cite{Kodama.H&Ishibashi2003A,Ishibashi.A&Kodama2003A}.

\subsection{Vector perturbation in higher dimensions}

Now, we show that even in higher dimensions, we can find a 
gauge-invariant master variable in terms of which the perturbation 
equations of each type can be reduced to a single second-order ODE 
of the Schr\"odinger type. We first treat vector perturbations.

Vector perturbations can be expanded in terms of vector harmonics on 
$\K^n$ satisfying
\Eq{
(\hat D\cdot\hat D + k_V^2)\VHB_i=0;\quad
      \hat D^i\VHB_i=0,
}
and harmonic tensors derived from them, 
\Eq{
\VHB_{ij}=-\frac{1}{2k_V}\left( \hat D_i\VHB_j +\hat D_j\VHB_i 
\right),
}
as
\Eq{
h_{ab}=0,\ 
h_{ai}=r{ f_a(y)}\VHB_i,\ 
h_{ij}=2r^2 { H_T(y)}\VHB_{ij}.
}
Here, in order to guarantee the completeness of the vector 
harmonics, we have assumed that $-\hat D\cdot\hat D$ is extended to 
a self-adjoint operator in the $L^2$-space of divergence-free vector 
fields on $\K^n$ by the Friedrichs extension. Note that $k_V^2\ge0$ 
in general, and for $K=1$, $k_V^2\ge 
n-1$\cite{Gibbons.G&Hartnoll2002,Kodama.H&Ishibashi2003Aa}.

For generic modes with $k_V^2\not=(n-1)K$, we can show that the combinations
\Eq{
F_a:=f_a +\frac{r}{k_V}D_a H_T
}
give a basis for gauge-invariant variables, where $D_a$ is the 
covariant derivative with respect to the metric $g_{ab}$ on the 
two-space $\N^2$. In contrast, for the exceptional modes with 
$k_V^2=(n-1)K$ ($K=0,1$), $\VHB_{ij}$ vanishes identically and $H_T$ 
is not defined. Hence, we obtain a smaller number of independent 
gauge-invariant variables. Since we can show that these exceptional 
modes correspond to adding rotation with some angular momentum and 
are not 
dynamical\cite{Kodama.H&Ishibashi2003A}, we do not discuss them in 
this article.
%

\begin{figure}
\centerline{
\begin{minipage}{5cm}
\includegraphics[height=6cm]{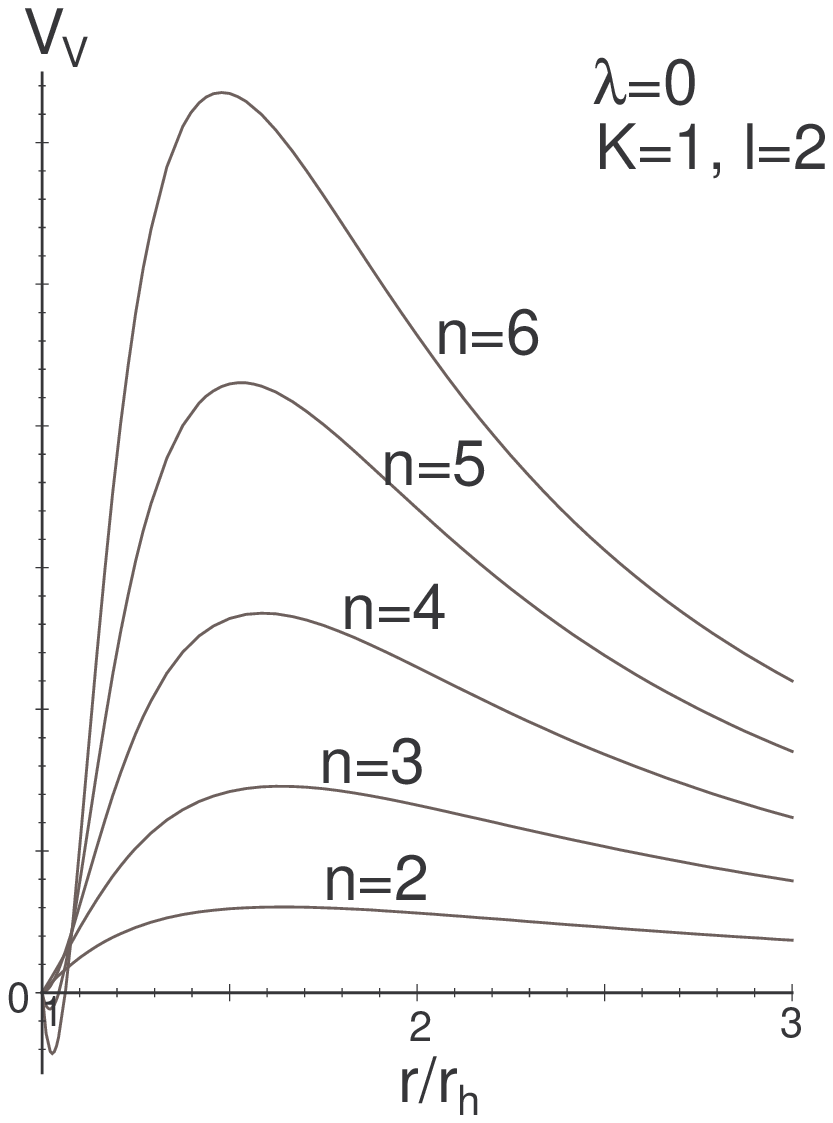}
\end{minipage}
\begin{minipage}{5cm}
\includegraphics[height=6cm]{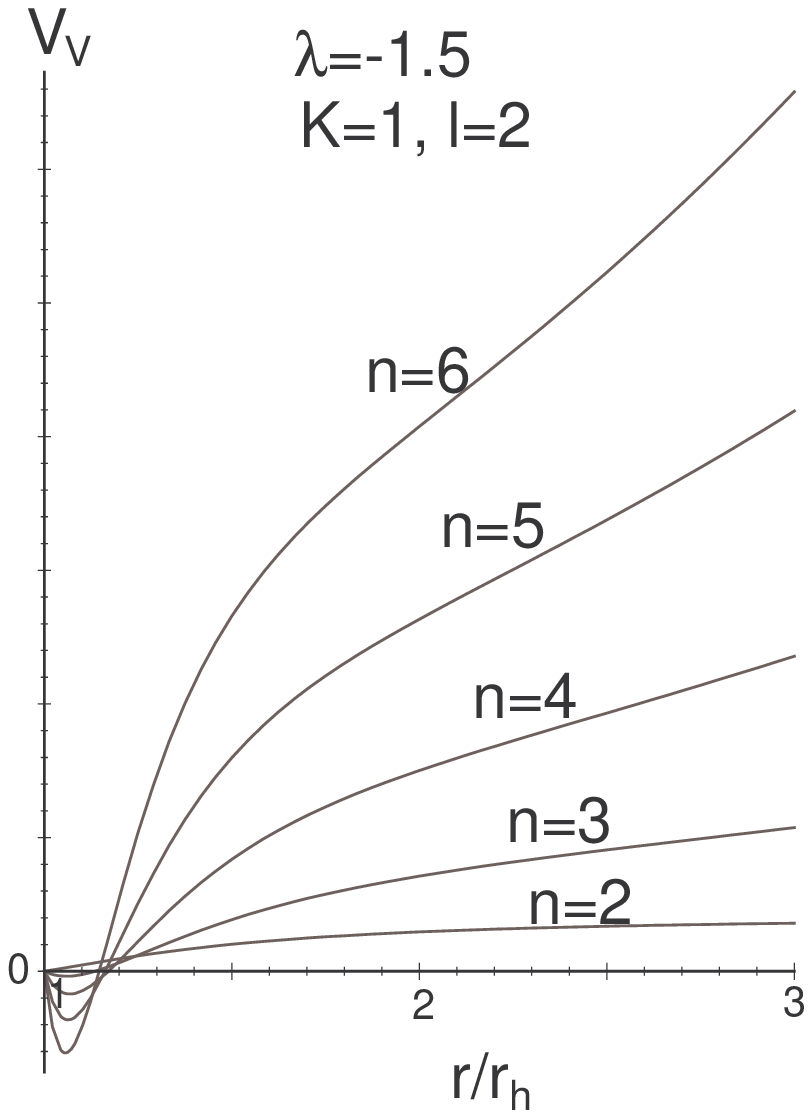}
\end{minipage}
}
\caption{Examples of $V_V$ for $\lambda=0$ and 
$\lambda<0$.}
\label{fig:V_V}
\end{figure}

The vacuum Einstein equations for a vector perturbation are expressed in terms of $F_a$ as
\Eqr{
&& D_a\left( r^{n+2}\epsilon^{bc}D_b(F_c/r) \right)
    -m_Vr^{n-1}\epsilon_{ab}F^b=0,\\
&& k_V D_a(r^{n-1}F^a)=0,
}
where $m_V=k_V^2-(n-1)K$. It is easy to see that these equations are equivalent to 
\Eqr{
&& r^{n-1}F^a=\epsilon^{ab}D_b\left( r^{n/2}\Phi \right),\\
&& -\frac{d^2\Phi}{dr_*^2}+V_V(r) \Phi=\omega^2\Phi,
}
where
\Eq{
V_V=\frac{f}{r^2}\left[ m_V+\frac{n(n+2)K}{4}
        -\frac{n(n-2)}{4}\lambda r^2 
        -\frac{3n^2M}{2r^{n-1}} \right].
}
%

Although the effective potential $V_V$ is not positive definite for 
large $n$, as is shown in Fig.\ref{fig:V_V}, we can show that there 
exists no unstable mode, i.e., no $L^2$-normalizable eigenfunction 
$\Phi$ with $\omega^2<0$ by the following argument, which we call 
{\it the $S$-deformation}.

Let $I$ be the range of $r_*$ corresponding to the regular region outside of the horizon, $-\infty <r_*<r_{*\infty}$. Here, $r_{*\infty}$ is $+\infty$ for $\lambda\ge0$, but is finite for $\lambda<0$.  Then, 
in the space $C_0^\infty(I)$ of smooth functions with compact support, 
the operator
\Eq{
A:=-\frac{d^2}{dr_*^2} + V(r)
}
is symmetric. We assume that $A$ is extended to a self-adjoint operator in  $L^2(I)$ by the Friedrichs extension. Then, { the lower bound for the spectrum of $A$ coincides with the lower bound of } 
\Eq{
\omega^2=(\Phi,A\Phi)/(\Phi,\Phi),\quad \Phi\in C^\infty_0(I).
}
Here, for any regular function $S(r)$ on $I$ and $\Phi\in 
C_0^\infty(I)$, a partial integration yields
\Eq{
(\Phi,A\Phi)=\int dr_*\left[ \left|\tilde D\Phi\right|^2 
             + \tilde V|\Phi|^2\right]
}
where
\Eqr{
&& \tilde D=\frac{d}{dr_*} + S(r),\\
&& \tilde V=V + f\frac{dS}{dr} -S^2.
}
Hence, if we can show that $\tilde V$ is non-negative for an 
appropriate $S$, we can conclude that $\omega^2\ge0$.

If we apply this method to a vector perturbation using 
$S=nf/(2r)$, we obtain
\Eq{
\tilde V_V=m_V \frac{f}{r^2}.
}
Since $m_V=k_V^2-(n-1)K$ is always non-negative, this implies that 
the black hole is stable against the vector perturbation in any 
spacetime dimension $D=n+2\ge4$, irrespective of the values of $K$ 
and $\lambda$.

\subsection{Scalar perturbation in higher dimensions}

Next, we consider scalar perturbations. Scalar perturbations can be 
expanded in terms of scalar harmonics on $\K^n$ satisfying
\Eq{
\left( \hat\triangle +k^2 \right)\SHB=0,
}
and harmonic vectors and tensors derived from them,
\Eqr{
&& \SHB_i :=-\frac{1}{k} \hat D_i\SHB,\\
&& \SHB_{ij} :=\frac{1}{k^2}\hat D_i\hat D_j\SHB
             +\frac{1}{n}\gamma_{ij}\SHB;\quad \SHB^i_i=0,
}
as
\Eq{
h_{ab}={ f_{ab}}\SHB,\quad 
h_{ai}=r{ f_a} \SHB_i,\quad 
h_{ij}=2r^2({ H_L}\gamma_{ij}\SHB +{ H_T}\SHB_{ij}).
}

Some comments are in order. First, since we are assuming that $\K^n$ 
is complete, $-\hat \triangle=-\hat D\cdot\hat D$ has a unique 
self-adjoint extension in $L^2(\K^n)$, which coincides with the 
Friedrichs extension and is non-negative. Second, we do not consider 
the zero modes $(k^2=0)$, by assuming that such a mode corresponds 
to a variation of the parameters of the background solution. This 
assumption is satisfied when $\K^n$ is compact and closed. Third, 
for $K=1$, the second smallest eigenvalue is $k^2=n$. For this 
eigenvalue, $\SHB_{ij}$ vanishes. Since we can show that such modes 
are pure gauge\cite{Kodama.H&Ishibashi2003A}, we do not consider 
them in this article.

From gauge transformation laws of the perturbation variables, we 
find that the following  combinations can be used as a basis for 
gauge-invariant variables\cite{Kodama.H&Ishibashi&Seto2000}:
\Eq{
F:=H_L +\frac{1}{n}H_T + \frac{D_ar}{r}X^a,\quad
F_{ab}:=f_{ab} +D_a X_b + D_b X_a,
}
where
\Eq{
X_a := \frac{r}{k}\left( f_a +\frac{r}{k}D_a H_T\right).
}
%

In terms of the gauge-invariant variable $\Phi$ constructed from 
these, 
\Eqr{
&& \Phi=\frac{nr^{n/2}}{H}\left( 2F+\frac{F^r_t}{i\omega r} 
\right);\\
&& H=m+\frac{n(n+1)M}{r^{n-1}},\quad  m=k^2-nK,
}
the Einstein equations are reduced to the single equation%
\cite{Kodama.H&Ishibashi2003A}
\Eq{
 -\frac{d^2\Phi}{dr_*^2}+V_S(r) \Phi=\omega^2\Phi.
}
The effective potential $V_S$ is given by
\Eq{
V_S(r)=\frac{f U(r)}{16r^2H^2}
}
with
\Eqr{
&U(r) =
 &-\left[n^3(n+2)(n+1)^2x^2-12n^2(n+1)(n-2)mx\right.
 \notag\\
&& \left. \qquad +4(n-2)(n-4)m^2\right] \lambda r^2  +n^4(n+1)^2x^3 \notag\\
&& +n(n+1)\left[4(2n^2-3n+4)m+n(n-2)(n-4)(n+1)K\right]x^2
   \notag\\
&& -12n\left[(n-4)m+n(n+1)(n-2)K \right]mx \notag\\
&&   +16m^3+4Kn(n+2)m^2,
\label{Q:NonStatic}
}
where $x=2M/r^{n-1}$.

\begin{figure}
\centerline{
\begin{minipage}{5cm}
\includegraphics[height=6cm]{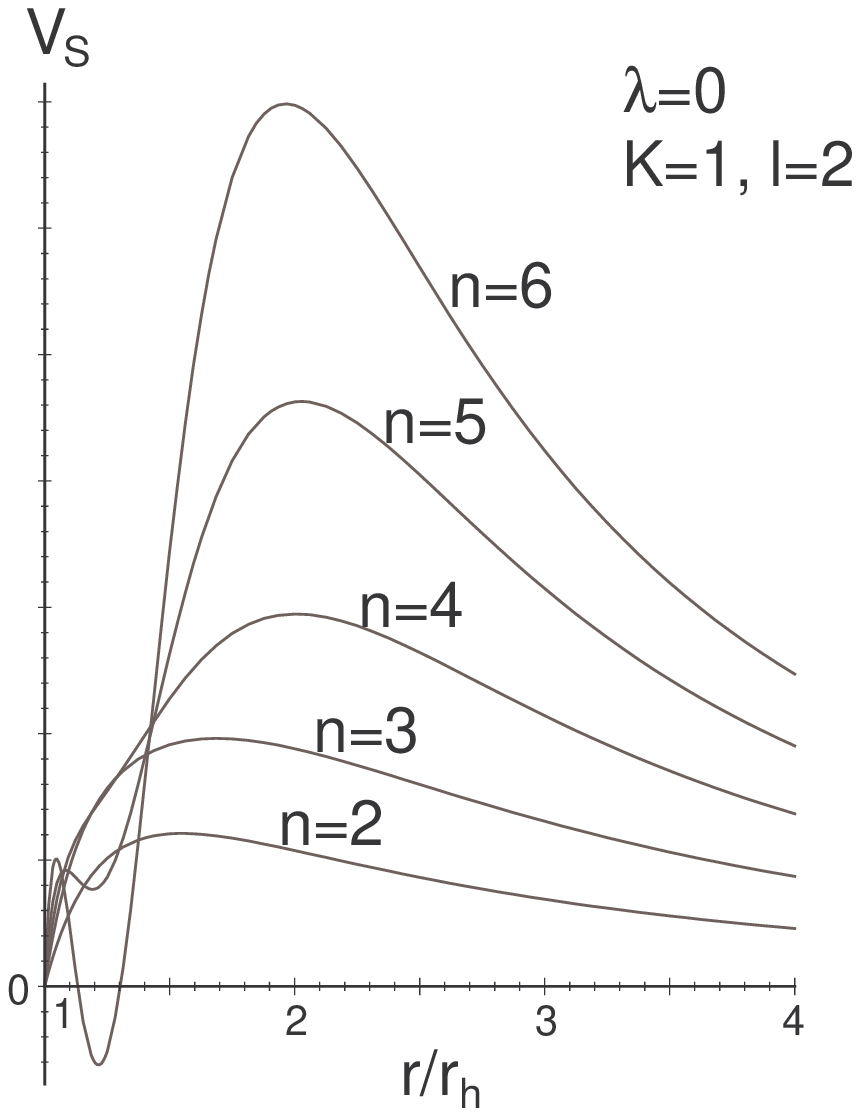}
\end{minipage}
\begin{minipage}{5cm}
\includegraphics[height=6cm]{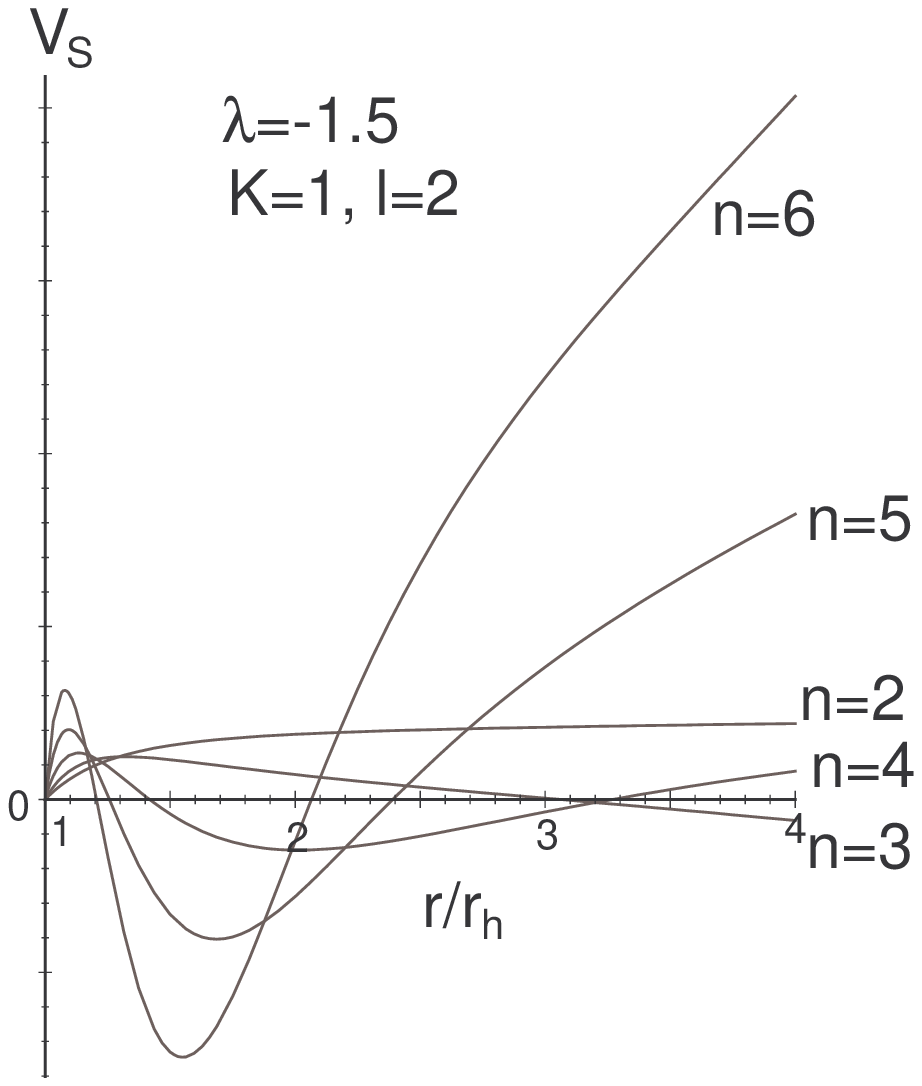}
\end{minipage}
}
\caption{Examples of $V_S$ for $\lambda=0$ and $\lambda<0$.}
\label{fig:V_S}
\end{figure}

We can easily see that for $n=2$ and $K=1$, $V_S$ is positive 
definite. Hence, 4D spherically symmetric neutral black holes are 
perturbatively stable, irrespective of the value of the cosmological 
constant. In contrast, $V_S$ is not positive definite for larger $n$ 
in general, as illustrated in Fig.\ref{fig:V_S}. Nevertheless, for 
$K=1$ and $\lambda=0$, neutral black holes are perturbatively stable 
in any spacetime dimensions, because the $S$-deformation of $V_S$ 
by
\Eq{
S=\frac{f}{h}\frac{dh}{dr};\quad
h=r^{n/2+l-1}H
}
yields
\Eq{
\tilde V_S=\frac{lx f}{2r^2H}
 \left[ ln(n+1)x+2(l-1)\left\{n^2+n(3l-2)+(l-1)^2\right\} \right]>0.
}
%

\subsection{Tensor perturbation}

In the case in which $\K^n$ is maximally symmetric, the only 
non-trivial tensor on $\K^n$ is the metric $\gamma_{ij}$. Hence, 
only the Laplacian appears as the derivative operator in the 
perturbation equation, and by the harmonic expansion with respect to 
$\hat D\cdot\hat D$, the Einstein equations reduce to equations on 
$\N^2$ for any type of perturbation. In contrast, when $\K^n$ is a 
generic Einstein space, the Weyl curvature couples tensor 
perturbations, and such a reduction cannot be achieved by the 
harmonic expansion. However, since the covariant derivatives and the 
Weyl curvature appear in the Einstein equations for tensor 
perturbations as the combination called  the Lichnerowicz operator%
\cite{Gibbons.G&Hartnoll2002}
\Eq{
\hat \triangle_L h_{ij} :
=-\hat D\cdot\hat D h_{ij}-2\hat R_{ikjl}h^{kl}+2(n-1)K h_{ij},
}
we can utilise the eigentensors of $\hat \triangle_L$,
\Eq{
\hat \triangle_L \THB_{ij}=\lambda_L \THB_{ij};\quad
\THB^i_i=0,\quad \hat D^j\THB_{ij}=0.
}
Tensor perturbations can be expanded in terms of these eigentensors as 
\Eq{
h_{ab}=0,\quad h_{ai}=0,\quad h_{ij}=2r^2{ H_T(y)}\THB_{ij}.
}
%

$H_T$ is gauge-invariant by itself, and the Einstein equations are reduced to
\Eq{
\Box H_T +\frac{n}{r}Dr\cdot 
DH_T-\frac{\lambda_L-2(n-1)K}{r^2}H_T=0.
}
In terms of 
\Eq{
\Phi= r^{n/2}H_T,
}
this equation is rewritten in the canonical form
\Eq{
 -\frac{d^2\Phi}{dr_*^2}+V_T(r) \Phi=\omega^2\Phi,
\label{MasterEq:Tensor}
}
where
\Eq{
V_T=\frac{f}{r^2}\left[\lambda_L-2(n-1)K+\frac{nrf'}{2}
          +\frac{n(n-2)f}{4}\right].
\label{V_T}
}
Note that this equation holds for any $f(r)$, if the background 
metric takes the form \eqref{BG:metric}.
 
For $f(r)$ given by \eqref{f:neutral}, \eqref{V_T} can be written as
\Eq{
V_T=\frac{f}{r^2}  
   \left[
      \frac{n(n+2)}{4}f+\frac{n(n+1)M}{r^{n-1}}+\lambda_L-(3n-2)K 
         \right]. 
}
When $\K^n$ is maximally symmetric, $\lambda_L$ is related to the 
eigenvalue $k_T^2$ of $-\hat D\cdot\hat D$ by $\lambda_L=k_T^2+2nK$, 
and $\lambda_L-(3n-2)K=(l-1)(l+n)>0$ for $K=1$. Hence, $V_T>0$ and 
the black hole is stable against the tensor perturbation. 

In contrast, when $\K^n$ is a generic Einstein space, no general 
lower bound for $\lambda_L$ is know, and generalised static black 
holes can be unstable against a tensor 
perturbation\cite{Gibbons.G&Hartnoll2002}.

\section{Extension to Charged Black Holes }


The extension of the formulations by Regge-Wheeler and Zerilli to 
Reissner-Nordstr\"om black holes was done by Moncrief, Zerilli and 
Chandrasekhar\cite{Moncrief.V1974,Zerilli.F1974,Chandrasekhar.S1983B},
 and used to prove their stability. In the charged black hole case, 
we have to take into account perturbations of electromagnetic 
fields, which couple metric perturbations non-trivially. However, by 
taking appropriate linear combinations of variables describing these 
perturbations, the perturbation equations for the Einstein-Maxwell 
system can be reduced to decoupled master equations, each of which 
is a single second-order ODE of the Schr\"odinger equation type. 
Now, we show that such a reduction is also possible in the case of 
higher-dimensional static black holes with charge, and discuss their 
stability\cite{Kodama.H&Ishibashi2003Aa}.

We assume that an background spacetime is again locally written as  
$\M^{n+2}\approx \N^2\times \K^n$, and the metric is given by 
\eqref{BG:metric} with \eqref{StaticOrbitSpace}. We further assume 
that the electromagnetic tensor $\F_{\mu\nu}$ for the background EM 
field takes the form
\Eq{
\F =\frac{1}{2}E_0 \epsilon_{ab}dy^a \wedge dy^b.
}
Then, the Maxwell equations and the Einstein equations determine $f(r)$ and $E_0(y)$ as
\Eqr{
&& f(r)=K-\lambda y^2-\frac{2M}{r^{n-1}}+\frac{Q^2}{r^{2n-2}},\\
&& E_0=\frac{q}{r^{n}};\quad 
   Q^2=\frac{\kappa^2q^2}{n(n-1)}.
}
%

\subsection{Tensor perturbation}

Since the EM field does not couple tensor perturbations of the 
metric, the master equation for the charged case is again given by 
\eqref{MasterEq:Tensor} with \eqref{V_T}. If we apply the 
$S$-deformation with $S=-nf/(2r)$ to $V_T$, we obtain 
\Eq{
\tilde V_T=\frac{f}{r^2}\left[ \lambda_L-2(n-1)K \right].
}
Hence, a generalised static black hole with charge is still stable 
against tensor perturbations, if $\lambda_L\ge 2(n-1)K$, or  
$k_T^2\ge -2K$ when $\K^n$ is maximally symmetric. This condition is 
satisfied by black holes with flat or spherical horizons.  However, 
the case $K=-1$ ($\lambda<0$) may be unstable even if $\K^n$ is 
maximally symmetric.

\subsection{Vector perturbation}

From the Maxwell equation $\nabla_{[\mu}\delta \F_{\nu\lambda]}=0$, 
a vector perturbation of the EM field is expanded in terms of the 
vector harmonics as
\Eq{
\delta \F_{ab}=0,\ 
\delta \F_{ai}={ D_a\A}\VHB_i,\ 
\delta \F_{ij}={ \A} \left(\hat D_i\VHB_j-\hat D_j\VHB_i\right).
}
Hence, a vector perturbation of the EM field is described by the 
single variable $\A(y)$, which is gauge-invariant by itself. For 
generic modes $m_V\equiv k_V^2-(n-1)K\not=0$, in terms of $\A$ and 
$F_a$, the Einstein equations are written as
\Eqr{
&& r^{n-1}F^a=\epsilon^{ab}D_b\Omega,\\
&&  r^nD_a\left( \frac{1}{r^n}D^a\Omega \right)
  -\frac{m_V}{r^2}\Omega 
  =\frac{2\kappa^2 q}{r^2}\A,
}
and the Maxwell equation $\delta(\nabla_\nu\F^{\mu\nu})=0$ are 
written as
\Eq{
\frac{1}{r^{n-2}}D_a(r^{n-2}D^a\A)-\frac{1}{r^2}
 \left(k_V^2+(n-1)K + \frac{2n(n-1)Q^2}{r^{2n-2}}\right)\A 
=\frac{qm_V}{r^{2n}}\Omega.
}
%

As shown in \cite{Kodama.H&Ishibashi2003Aa}, by taking linear 
combinations $\Phi_\pm =a_\pm r^{-n/2}\Omega + b_\pm r^{n/2-1}\A$ 
with some constants $a_\pm$ and $b_\pm$, these perturbation 
equations for the Einstein-Maxwell system are transformed to the two 
decoupled equations
\Eq{
\frac{d^2\Phi_\pm}{dr_*^2}+V_{V\pm}(r)\Phi_\pm=\omega^2\Phi_\pm.
}
The effective potentials $V_{V\pm}$ are given by
\Eq{
V_{V\pm}=\frac{f}{r^2}\left[k_V^2 +\frac{(n^2-2n+4)K}{4}
   -\frac{n(n-2)}{4}\lambda r^2+\frac{n(5n-2)Q^2}{4r^{2n-2}} 
   +\frac{\mu_\pm}{r^{n-1}}\right],
}
where
\Eq{
\mu_\pm:=-\frac{n^2+2}{2}M \pm \Delta;\quad
 \Delta:=\left[(n^2-1)^2M^2+2n(n-1)m_V Q^2\right]^{1/2}.
}

In the limit $Q=0$, $\Phi_+$ and $\Phi_-$ coincide with $\A$ and 
$\Omega$, respectively. Hence, $\Phi_+$ and $\Phi_-$ represent the 
EM mode and the gravitational mode of the perturbation, respectively.

By the $S$-deformation with $S=\frac{nf}{2r}$, $V_{V\pm}$ are 
transformed to 
\Eq{
\tilde V_{V\pm} =\frac{f}{r^2}\left[ m_V
   +\frac{(n^2-1)M\pm\Delta}{r^{n-1}} \right].
}
Hence, $\tilde V_{V+}$ is always positive. However, since  
$\Delta>(n^2-1)M$ for $m_V>0$, $\tilde V_{V-}$ can become negative. 

By examining the behaviour of $\tilde V_{V-}$, we find that it is 
positive outside the horizon for $K=0$ and $K=1$, while it becomes 
negative near the horizon for $K=-1$ if $\lambda$ is sufficiently 
close to $\lambda_{c-}$, where $\lambda_{c-}$ is the lower bound on 
$\lambda$ for a fixed value of $Q$ when the spacetime contains a 
regular black hole\cite{Kodama.H&Ishibashi2003Aa}. Hence, we can 
conclude that a charged black hole with a flat or spherical horizon 
is stable against vector perturbations. However, we cannot prove the 
stability of a charged black hole with hyperbolic horizon by this 
simple method.

\subsection{Scalar perturbation}

A scalar perturbation of the EM field is expressed as
\Eqr{
&& \delta \F_{ab}+D_c(E_0X^c)\epsilon_{ab}\SHB
 =\frac{\epsilon_{ab}}{r^n}\left(-k^2{ \A}+\frac{q}{2}(F^c_c-2nF) \right)
  \SHB,\\
&& \delta\F_{ai}-kE_0\epsilon_{ab}X^b\SHB_i
  =\frac{k}{r^{n-2}}\epsilon_{ab}{ D^a\A}\,\SHB_i,\\
&& \delta\F_{ij}=0.
}
The expansion coefficient $\A$ is gauge invariant and obeys the 
equation
\Eq{
r^{n-2}D_a\left( \frac{D^a\A}{r^{n-2}} \right) -\frac{k^2}{r^2}\A
  =    -\frac{q}{2r^2}(F^c_c-2nF).
}
The Einstein equations for the scalar perturbation are equivalent to
\Eq{
 fD_a D^a \Phi - V_S(r)\Phi= \frac{f P_{S1}(r)}{r^{n/2}H^2}\kappa^2E_0\A,
}
where $P_{S1}(r)$ is a function of $r$, and 
\Eqr{
&& \Phi=\frac{nr^{n/2}}{H} \left( \frac{F^r_t}{i\omega r}+2F \right),\\
&& H=m +\frac{n(n+1)M}{r^{n-1}}-\frac{n^2Q^2}{r^{2n-2}};\quad m=k^2-nK.
}
%

In terms of the variables $\Phi_\pm$ written as linear combinations 
of $\Phi$ and $\A$, these perturbation equations can be rewritten as 
the two decoupled equations
\Eq{
\frac{d^2\Phi_\pm}{dr_*^2}+V_{S\pm}(r)\Phi_\pm=\omega^2\Phi_\pm.
}
Here, the effective potentials $V_{S\pm}$ are given by
\Eq{
V_{S\pm} =\frac{fU_\pm}{64r^2H_\pm^2}.
}
where in terms of the positive constant $\delta$ defined by
\Eq{
Q^2=(n+1)^2M^2\delta(1+m\delta),
}
$H_\pm$ are expressed as
\Eq{
H_+=1-\frac{n(n+1)\delta M}{r^{n-1}},\
H_-=m+\frac{n(n+1)(1+m\delta)M}{r^{n-1}},
}
and $U_\pm$ are expressed in terms of $x=\frac{2M}{r^{n-1}}$ as
\Eqr{
& U_+ =
& \left[-4 n^3 (n+2) (n+1)^2 \delta^2 x^2-48 n^2 (n+1) (n-2) \delta x
\right.\notag\\
&&\left.   -16 (n-2) (n-4)\right] \lambda r^2 \notag\\
&&  -\delta^3 n^3 (3 n-2) (n+1)^4 (1+m \delta) x^4
\notag\\
&&   +4 \delta^2 n^2 (n+1)^2 
   \left\{(n+1)(3n-2) m \delta+4 n^2+n-2\right\} x^3
\notag\\   
&&   +4 \delta (n+1)\left\{
   (n-2) (n-4) (n+1) (m+n^2 K) \delta \right.\notag\\
&& \qquad\qquad \left.-7 n^3+7 n^2-14 n+8
   \right\}x^2
\notag\\   
&&  + \left\{16 (n+1) \left(-4 m+3 n^2(n-2) K\right) \delta
     -16 (3 n-2) (n-2) \right\}x
\notag\\   
&&    +64 m+16 n(n+2) K.
}
\Eqr{
& U_- =
  & \left[-4 n^3 (n+2) (n+1)^2 (1+m \delta)^2 x^2 \right.\notag\\
&&\left.      +48 n^2 (n+1) (n-2) m (1+m \delta) x 
    -16 (n-2) (n-4) m^2\right] \lambda r^2 \notag\\
&&     -n^3 (3 n-2) (n+1)^4 \delta (1+m \delta)^3 x^4
\notag\\
&& -4 n^2 (n+1)^2 (1+m \delta)^2 
     \left\{(n+1)(3 n-2) m \delta-n^2\right\} x^3
\notag\\  
&&  +4 (n+1) (1+m \delta)\left\{ m (n-2) (n-4) (n+1) (m+n^2 K) \delta
  \right. \notag\\
&& \left. \quad  +4 n (2 n^2-3 n+4) m+n^2 (n-2) (n-4) (n+1)K \right\}x^2
\notag\\
&&  -16m \left\{ (n+1) m \left(-4 m+3 n^2(n-2) K\right) \delta
\right.\notag\\
&&\left.  +3 n (n-4) m+3 n^2 (n+1) (n-2)K \right\}x
\notag\\
&&      +64 m^3+16 n(n+2)m^2 K.
}
%

\begin{figure}
\centerline{
\begin{minipage}{5cm}
\centerline{\includegraphics[height=6cm]{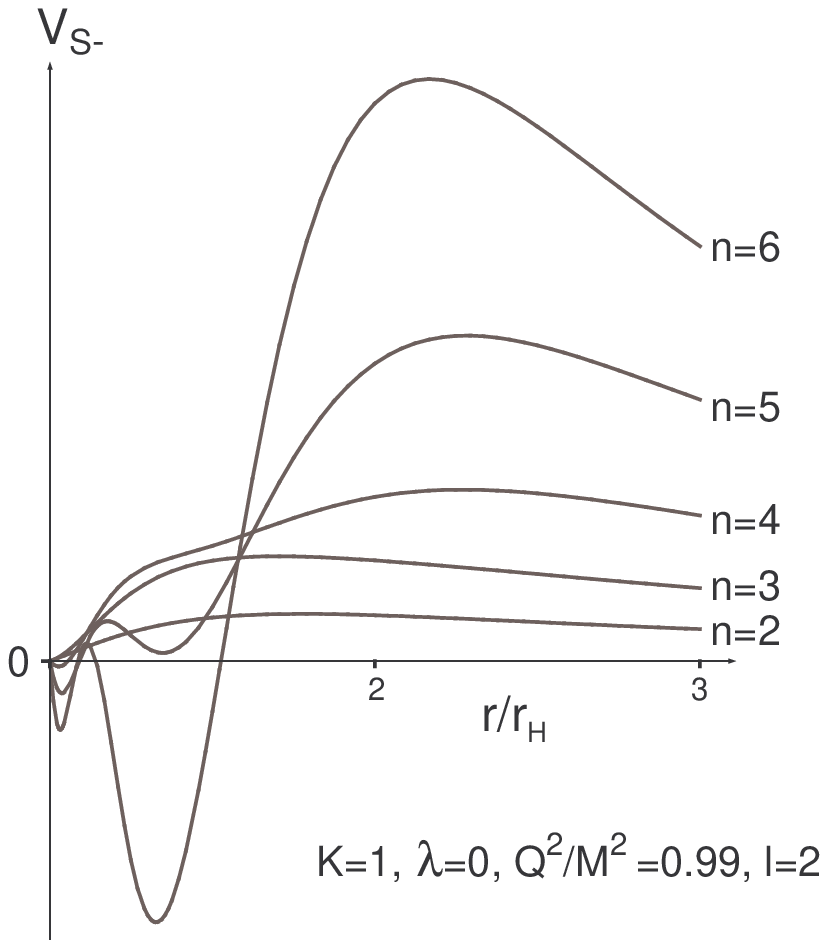}}
\end{minipage}
\begin{minipage}{5cm}
\centerline{\includegraphics[height=6cm]{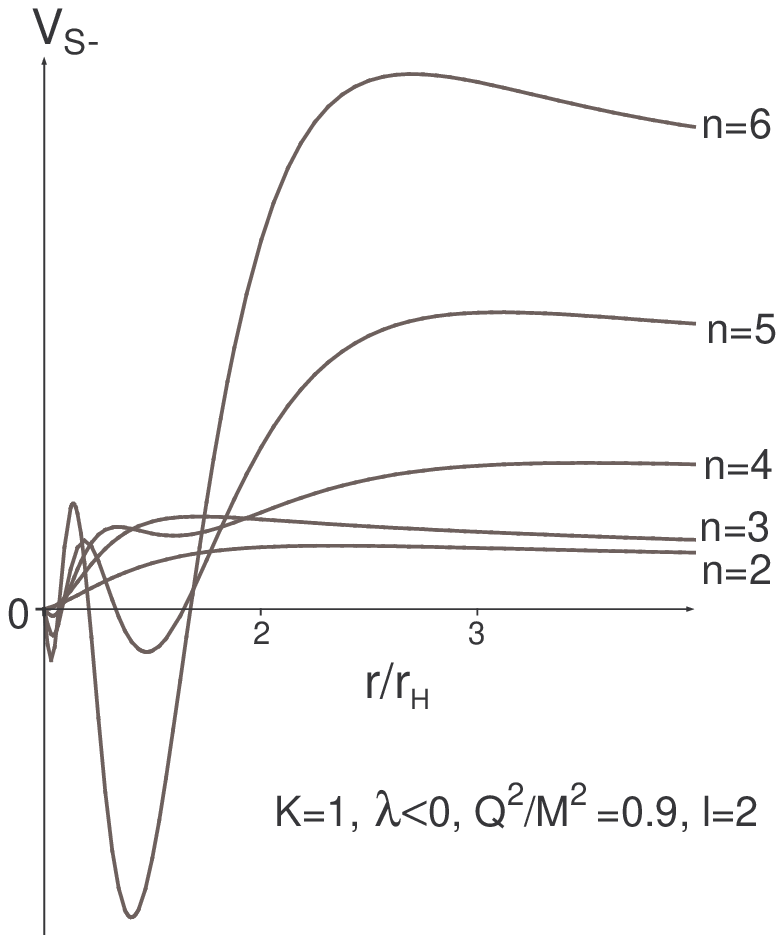}}
\end{minipage}
}
\caption{Examples of $V_{S-}$ for $\lambda=0$ and 
$\lambda<0$.}\label{fig:VSminus}
\end{figure}

By the $S$-deformation with 
\Eq{
S=\frac{f}{h}\frac{dh}{dr};\quad
h=r^{n/2-1}H_+,
}
the effective potential $V_{S+}$ for $\Phi_+$ is transformed to 
\Eq{
\tilde V_{S+}=\frac{k^2 f }{2r^2 H_+}
  \left[ (n-2)(n+1)\delta x + 2\right].
}
Since this is positive definite, the EM mode is always stable.

By the similar $S$-deformation
\Eq{
S=\frac{f}{h}\frac{dh}{dr};\quad
h=r^{n/2-1}H_-,
}
the effective potential $V_{S-}$ for $\Phi_-$ is transformed to 
\Eq{
\tilde V_{S-}=\frac{k^2f}{2r^2H_-}
  \left[ 2m-(n+1)(n-2)(1+m\delta)x \right].
}
Hence, the gravitational mode is also stable for $n=2$. 

However, for $n>2$, $\tilde V_{S-}$ becomes negative near the 
horizon in general (See Fig.\ref{fig:VSminus}). The $S$-deformation 
used in the neutral case does not work either, because it cannot 
remove a negative ditch of $V_{S-}$ produced by the charge near the 
horizon.  We have not been able to prove the stability for $n>2$ by 
this method, except for some special cases.

\section{Summary}

\begin{table}
\caption{Stability of static black holes.}\label{tbl:stability}
\centerline{\tiny
\begin{tabular}{l|l|c|c|c|c|c|c|}
\hline\hline
\multicolumn{2}{c|}{}& \multicolumn{2}{c|}{ Tensor}
  & \multicolumn{2}{c|}{ Vector}& \multicolumn{2}{c|}{ Scalar}\\
\cline{3-8}
\multicolumn{2}{c|}{}&$Q=0$ & $Q\not=0$ &$Q=0$ & $Q\not=0$ 
&$Q=0$ & $Q\not=0$ \\
\hline
$K=1$& $\lambda=0$ & OK & OK & OK & OK 
     & OK 
     & $\begin{array}{l} 
         D=4,5\ \text{OK} \\ D\ge6\ \text{?} 
       \end{array}$
     \\
\cline{2-8}
     &$\lambda>0$ & OK & OK & OK & OK 
     & $\begin{array}{l} 
         D\le6\ \text{OK} \\ D\ge7\ \text{?} 
        \end{array}$
     & $\begin{array}{l}
         D=4,5\ \text{OK} \\ D\ge6\ \text{?} 
        \end{array}$
     \\
\cline{2-8}
     &$\lambda<0$ & OK & OK & OK & OK 
     &  $\begin{array}{l}
          D=4\ \text{OK} \\ D\ge5\ \text{?} 
         \end{array}$
     &  $\begin{array}{l}
          D=4\ \text{OK} \\ D\ge5\ \text{?} 
         \end{array}$
     \\
\hline
$K=0$ &$\lambda<0$ & OK & OK & OK & OK 
     & $\begin{array}{l}
         D=4\ \text{OK} \\ D\ge5\ \text{?} 
        \end{array}$
     & $\begin{array}{l}
         D=4\ \text{OK} \\ D\ge5\ \text{?} 
       \end{array}$ 
     \\
\hline
$K=-1$ &$\lambda<0$ & OK & ? & OK & ? 
     & $\begin{array}{l}
         D=4\ \text{OK} \\ D\ge5\ \text{?} 
        \end{array}$
     & $\begin{array}{l}
         D=4\ \text{OK} \\ D\ge5\ \text{?} 
        \end{array}$
     \\
\hline
\end{tabular}
}
\end{table}

In this article, we have discussed the stability of generalised 
static black holes with and without charge in higher dimensions. The 
results of our analysis for maximally symmetric black holes are 
summarised in Table \ref{tbl:stability}\footnote{In this table, the 
results for vector and scalar perturbations are also valid for 
generalised static black holes, except for the case $K=1$, 
$\lambda>0$ and $Q=0$, for which the stability has been proved for 
$D\le5$ in spherically symmetric case.}, 

In particular for the case of $\Lambda=0$, we have established the 
stability of a spherically symmetric black hole for arbitrary 
dimensions, and that of of a spherically symmetric black hole with 
charge for $D=4$ and $D=5$. We have also proved the stability of 4D 
and 5D dS-Schwarzschild and 4D AdS-Schwarzschild black holes with 
charge. Although we have not succeeded in proving the stability of 
black holes for the other cases, there is no strong reason to 
suspect the instability of such black holes because the negative 
ditch of the effective potential is not so deep.

Finally, we would like to point out that the master equations used 
in this article for the stability analysis can be also used in other 
problems related to black holes, such as perturbative analyses of 
the uniqueness of asymptotically de Sitter or anti-de Sitter static 
black holes in four and higher dimensions, the precise determination 
of the frequencies of quasinormal modes and the greybody factor of 
the Hawking radiation, and the estimation of the gravitational wave 
emission rate from mini black holes, which might be produced in 
colliders and cosmic ray events\footnote{Explicit expressions for 
the source terms of the master equations are given in 
Ref.\cite{Kodama.H&Ishibashi2003Aa}}.

\section*{Acknowledgements}

AI is a JSPS fellow, and HK is supported by the JSPS grant No. 
15540267.


\end{document}